\documentclass[twocolumn,superscriptaddress,groupedaddress]{revtex4}  
\usepackage{graphicx}  
\usepackage{xcolor}
\usepackage{dcolumn}   
\usepackage{bm}        
\usepackage{amssymb}   
\usepackage{amsmath}   
\usepackage{multirow}
\usepackage{url}       
\usepackage[normalem]{ulem} 
\usepackage{xspace}
\usepackage[final]{pdfpages}
\newcommand{\SD}[1]{\ensuremath{S_{#1}}\xspace}

\begin{document}
\title{Methodology for determining the electronic thermal conductivity of metals via direct non-equilibrium ab initio molecular dynamics}

\author{Sheng-Ying Yue}
\affiliation{Aachen Institute for Advanced Study in Computational Engineering Science (AICES),RWTH Aachen University, 52062 Aachen, Germany}
\author{Xiaoliang Zhang}
\affiliation{Institute of Mineral Engineering, Division of Materials Science and Engineering, Faculty of Georesources and Materials Engineering, RWTH Aachen University, 52064 Aachen, Germany}
\author{Stephen Stackhouse}
\affiliation{School of Earth and Environment, University of Leeds, Leeds LS2 9JT, United Kingdom}
\author{Guangzhao Qin}
\affiliation{Institute of Mineral Engineering, Division of Materials Science and Engineering, Faculty of Georesources and Materials Engineering, RWTH Aachen University, 52064 Aachen, Germany}
\author{Edoardo Di Napoli}
\affiliation{Aachen Institute for Advanced Study in Computational Engineering Science (AICES),RWTH Aachen University, 52062 Aachen, Germany}
\affiliation{J\"ulich Supercomputing Centre, Forschungszentrum J\"ulich and JARA--HPC, 52425 J\"ulich, Germany}
\author{Ming Hu}\email{hum@ghi.rwth-aachen.de}
\affiliation{Aachen Institute for Advanced Study in Computational Engineering Science (AICES),RWTH Aachen University, 52062 Aachen, Germany}
\affiliation{Institute of Mineral Engineering, Division of Materials Science and Engineering, Faculty of Georesources and Materials Engineering,
RWTH Aachen University, 52064 Aachen, Germany}

\maketitle
\textbf{Many physical properties of metals can be understood in terms of the free electron model~\cite{kittel}, as proven by the Wiedemann-Franz law~\cite{William}. According to this model, electronic thermal conductivity ($\kappa_{el}$) can be inferred from the Boltzmann transport equation (BTE). However, the BTE does not perform well for some complex metals, such as Cu. Moreover, the BTE cannot clearly describe the origin of the thermal energy carried by electrons or how this energy is transported in metals. The charge distribution of conduction electrons in metals is known to reflect the electrostatic potential~(EP)~of the ion cores~\cite{kittel}. Based on this premise, we develop a new methodology for evaluating $\kappa_{el}$ by combining the free electron model and non-equilibrium ab initio molecular dynamics (NEAIMD) simulations. We demonstrate that the kinetic energy of thermally excited electrons originates from the energy of the spatial electrostatic potential oscillation (EPO), which is induced by the thermal motion of ion cores. This method directly predicts the $\kappa_{el}$ of pure metals with a high degree of accuracy.  }

\begin{figure*}
\includegraphics[width=0.8\linewidth,clip]{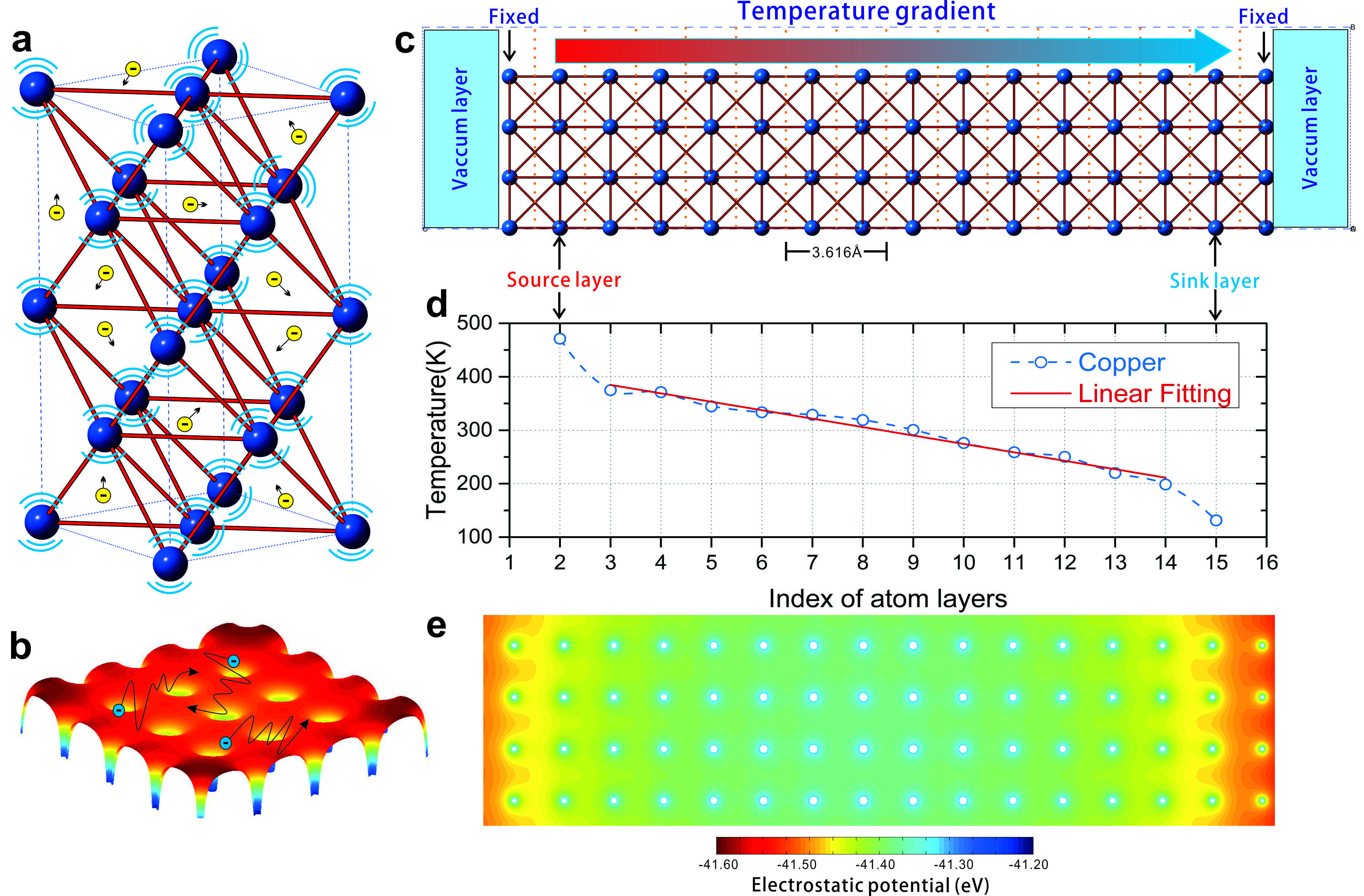}
\caption{Overview of the simulation model, temperature profile and EP field of copper. \textbf{a,~b,} Cartoons of free electrons in a metal moving in the vibrating lattice and EP field. \textbf{c,~d,} Model of copper used in NEAIMD simulations and the corresponding temperature profile. One unit cell length comprises two layers of atoms. We use fixed boundary conditions with the layers of fixed atoms and vacuum layers along the direction of $\nabla T$. Periodic boundary conditions are adopted in the other two dimensions. \textbf{e,} Theoretical EP field of a perfect copper structure (the test charge number is norm 1).}
\label{fig:1}
\end{figure*}

\begin{figure*}
\includegraphics[width=1.0\linewidth,clip]{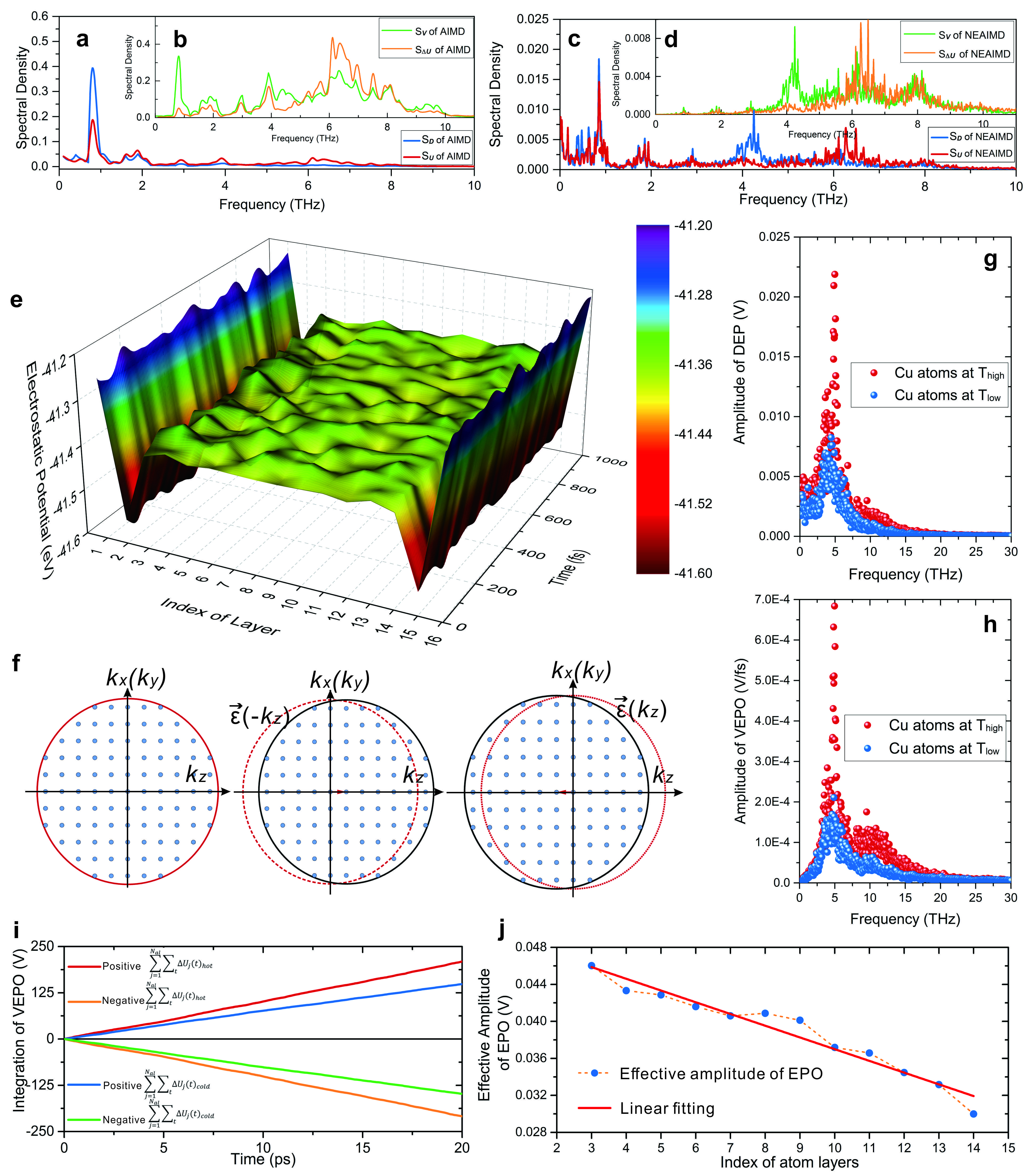}
\caption{\label{fig:2} Overview of the relationship between EPO at ion cores and lattice vibrations. \textbf{a,~c,} Blue and red lines, represent the spectral density of atom displacements ($\SD{D}$) and the EP displacement ($\SD{U}$) at specific ion cores. \textbf{b,~d,} Green and orange lines, represent the spectral density of atom velocities ($\SD{V}$) and the EPO velocities ($\SD{\Delta U}$), respectively, at specific ion cores. Data in \textbf{(a,~b)} are from a 10-ps equilibrium AIMD simulation of Al at 100.90\,K, whereas \textbf{(c,~d)} present the same physical quantities from a 70-ps NEAIMD simulation of Al at 299.46\,K. Data shown in \textbf{(e,~g-j)} are from a 20-ps NEAIMD simulation of Cu at 298.49\,K. \textbf{e,} The variation of EPO in space over time (the test charge number is 1).  \textbf{f,} Schematic of the whole Fermi sphere oscillation as local electric field vibration along the $\vec{z}$ direction.  \textbf{g,~h,} The direct fast Fourier transform (FFT) amplitudes of the displacement of EP (DEP, $U_{\rm ion}$) and VEPO ($\Delta U_{\rm ion}$) of ion cores at different temperatures. \textbf{i,} Integration of positive and negative VEPO ($\sum_{j=1}^{N_{al}}\sum_{t}\Delta U_{j}(t)$) at different temperatures. \textbf{j,} The effective amplitude of EPO ($\frac{1}{N_{al}}\sum_{j=1}^{N_{al}}\sqrt{\frac{1}{n_{steps}}\sum_{i=1}^{n_{steps}}(U_{ij}-\overline{U}_{j})^2}$, average root mean square (RMS)~\cite{John} of EPO, where $N_{al}$ is the atom number per layer) in atom layers along the $\vec{z}$ direction.}
\end{figure*}

\begin{figure*}
\includegraphics[width=0.9\linewidth,clip]{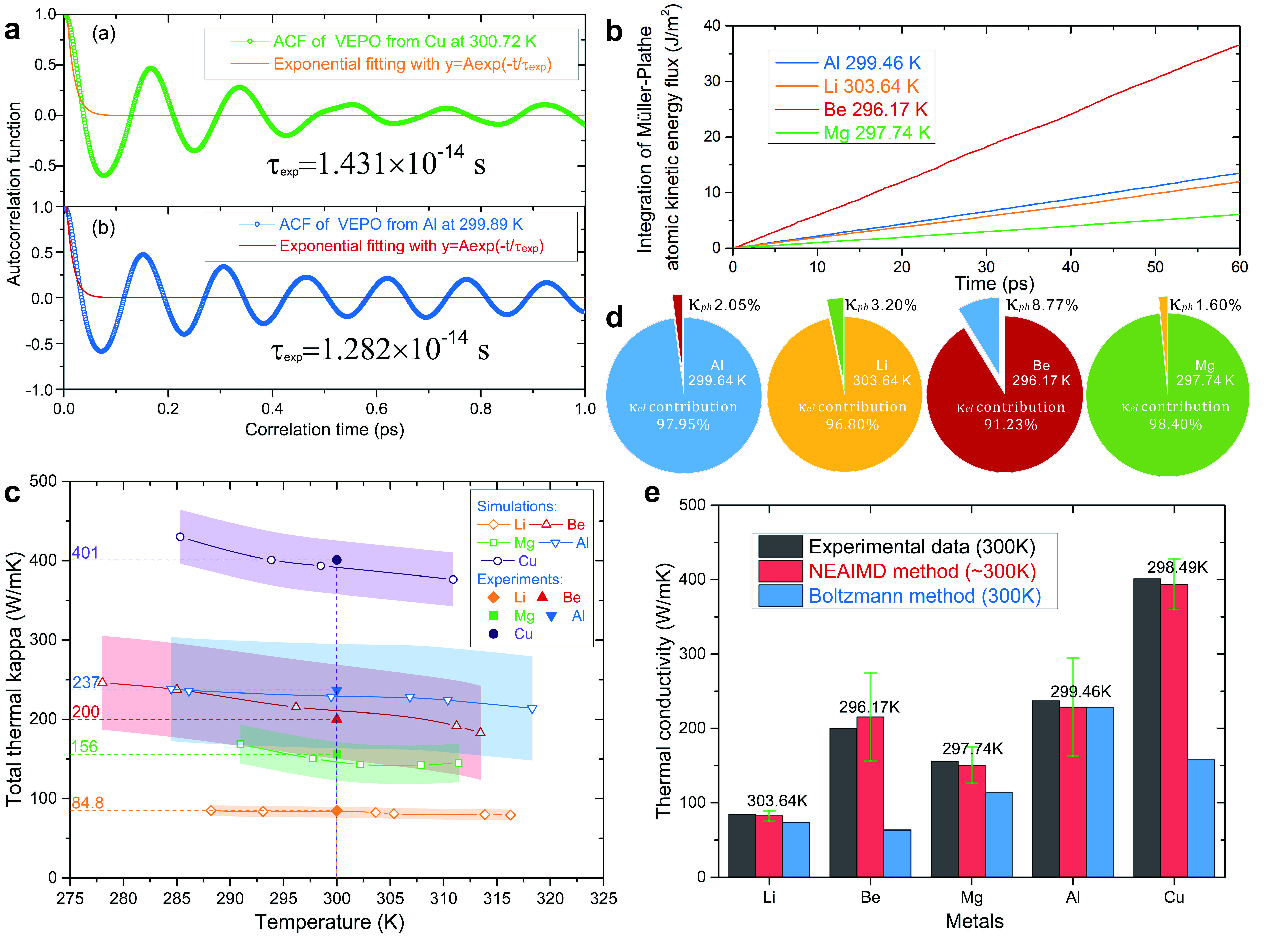}
\caption{\label{fig:3} NEAIMD-EPO simulation results for metals. \textbf{a,} Autocorrelation function of VEPO from Cu and Al NEAIMD simulations. The exponential fitting followed the function: $y=A\mathrm{exp}(-t/\tau_{\mathrm{exp}})$, where $\tau_{\mathrm{exp}}$ represents the exponential autocorrelation time of VEPO. \textbf{b,} Integration of the atomic kinetic energy flux with time based on the NEAIMD (M\"uller-Plathe) simulations of Al, Li, Be and Mg. \textbf{c,} Total thermal conductivities of metals from NEAIMD simulation (with error bars determined from the calculation of $\nabla T$ and $\frac{\partial\overline{U}_{EPO}(l)}{\partial N_{l}}$ along the heat-transport direction) and experimental data at 300\,K. \textbf{d,} Pie graphs showing the electronic and phononic contributions to the total thermal conductivity of Al, Li, Be, and Mg. \textbf{e,} Bar graph comparing the thermal conductivities of metals calculated using the NEAIMD method (with error bars), Boltzmann method and experimental data at 300\,K.}
\end{figure*}

$\kappa_{el}$ is one of the most important physical properties of metals. The analytical solution of $\kappa_{el}$ based on the BTE and free electron model can be expressed as
\begin{equation}
\kappa_{el}= \frac{\pi^2 n k_B^2 T \tau_{el}}{3m},
\end{equation}
where $n$ is the concentration of free electrons, $m$ is the electron mass, $k_{B}$ is the Boltzmann constant, $T$ is the system temperature and
$\tau_{el}$ is the collision time of free electrons, which is mainly determined by electron-electron, electron-hole and electron-phonon
scattering.  In principle, we can obtain an approximate value for $\tau_{el}$ from Matthiessen's rule. However, describing every scattering process involved in the
heat transfer by electrons of solid metals is too complicated. Recently, several methods have been reported for the evaluation of the electronic thermal conductivities of liquid-phase metals within the framework of density functional theory (DFT), such as ab initio molecular dynamics (AIMD) using the Kubo-Greenwood equation~\cite{Collins, Pozzo,Nico} and DFT plus dynamical mean-field theory (DFT+DMFT)~\cite{Zhang}.

In this paper, we develop a new methodology to describe the electronic heat-transport process in solid metals without explicitly addressing detailed scattering processes. From the second law of thermodynamics, we know that heat transfer in solids is driven by the temperature gradient $\nabla T$. Typically, temperature describes the thermal motion of atoms, and the vibrations of ions can lead to spatial EPO, as can be easily deduced from the mathematical expression for the total Hamiltonian of system. Local variation of the EP can drive the collective oscillation of free electrons, and those free electrons near the Fermi surface can be excited above the Fermi surface and obtain additional thermal kinetic energy with respect to $0\ \mathrm{K}$. These are called thermally excited electrons. Figs.~\ref{fig:1}(a,~b) show two cartoons describing how the thermally excited electrons move in the vibrational lattice and the local EP field. Higher temperatures, which induce larger and faster ionic vibrations, lead to stronger EPO. Thus, the thermally excited electrons in high-temperature regions have more kinetic energy than those in low-temperature regions. Once a stable distribution of the thermal kinetic energy of thermally excited electrons is established along the direction of $\nabla T$, then the heat flux carried by thermally excited electrons and $\kappa_{el}$ can be calculated.

To prove this conjecture and quantify $\kappa_{el}$, we performed NEAIMD simulations~\cite{Stephen,kaviany} by modifying the Vienna Ab initio Simulation Package (VASP)~\cite{vasp01,vasp02}. The atomic heat flux was realized using the M\"uller-Plathe algorithm~\cite{Muller-Plathe}, in which the kinetic energies of the atoms in the heat source and heat sink are exchanged (Supplementary Information (SI) Sec.~1). With sufficient simulation time, we can establish a stable temperature gradient in metals. Figs.~\ref{fig:1}(c,~d) present the Cu model and the corresponding temperature profile, respectively.
Simultaneously, we can calculate the spatial distribution and the dynamical evolution of the EP, which is expressed as
\begin{eqnarray}
U=\int U(r)\cdot\rho_{\mathrm{test}}\cdot(\vert r-R\vert)d^{3}r,
\end{eqnarray}
where the test charge $\rho_{\mathrm{test}}$ is the norm $1$, and $R$ represents the ion position. Fig.~\ref{fig:1}(e) shows the theoretical results of
the static distribution of the EP for a perfect Cu lattice. In the rest of this paper, we 1) illustrate the relationship between the spatial EPO and lattice vibrations, 2) demonstrate that the EPO provides additional kinetic energy to thermally excited electrons, and 3) show how to predict $\kappa_{el}$ within our theoretical framework.

To demonstrate the relationship between EPO and lattice vibrations, we analyse the data from our AIMD simulations using the power spectral density (PSD) method~\cite{Dennis,Storch}. For a stationary signal $x(t)$, the PSD is defined as
\begin{eqnarray}
S_{x}(f)=\int_{-\infty}^{\infty}R_{x}(\tau)e^{-2\pi if\tau}d\tau,
\end{eqnarray}  
where $R_{x}(\tau)=E[x(t)x(t+\tau)]$ is the autocorrelation function of $x(t)$~\cite{Dennis,Storch}, and $E[\ \cdot\ ]$ denotes the
expectation value. Here, we consider four signals from an AIMD simulation: atomic displacement $D_{\rm ion}$, atomic velocity $V_{\rm ion}$, EP displacement $U_{\rm ion}$, and velocity of EPO (VEPO) $\Delta U_{\rm ion}$; these are used to calculate their respective spectral densities $\SD{D}$, $\SD{V}$~\cite{Xiaoliang}, $\SD{U}$, $\SD{\Delta U}$ (SI Sec.~2). $\SD{D}$ and $\SD{V}$ reflect the frequency-dependent lattice vibrations at a specific
$T$. Analogously, $\SD{U}$ and $\SD{\Delta U}$ provide information regarding the EPO with respect to frequency. We show results for Al from a 10-ps equilibrium AIMD run at 100.90\,K (Figs.~\ref{fig:2}(a,~b)) and a 70-ps NEAIMD simulation at 299.46\,K (Figs.~\ref{fig:2}(c,~d)). Fig.~\ref{fig:2}(a) clearly shows that the locations of the density peaks of $\SD{D}$ and $\SD{U}$ are consistent, demonstrating that the EPO is mainly caused by the lattice vibration of ion cores. Fig.~\ref{fig:2}(b) confirms this relationship. Similar results are shown in Figs.~\ref{fig:2}(c,~d): for most of the frequency ranges, the peaks of $\SD{D}$ and $\SD{U}$, $\SD{V}$ and $\SD{\Delta U}$ are consistent with each other. However, for some specific frequencies in Figs.~\ref{fig:2}(c,~d) ($3.5\sim 5.0$ THz), some discrepancies exist in the peaks' magnitudes. This phenomenon may be attributable to the heat flux applied in the NEAIMD simulations. Nevertheless, Figs.~\ref{fig:2}($a\sim d$) provide unambiguous evidence that the EPO is directly induced by the lattice vibration of ions in metals.

To understand the dynamical evolution of spatial EPO intuitively, we present the representative case of Cu, calculated using NEAIMD at 298.49\,K, in Fig.~\ref{fig:2}(e). Little variation occurs in the local electronic field between neighbouring atom layers, and the directions of these local fields continually change with time. The variations of these local fields will drive the collective vibration of free electrons, as theoretically illustrated in Fig.~\ref{fig:2}(f). We see that only free electrons near the Fermi surface can be thermally excited. Because the direction of the local field continually changes with time, the vectors of the local momentum of the thermally excited electrons should also continually change with time. Therefore, for a sufficiently long statistical time average, no net electric current should arise during the thermal transport process of metals. This is consistent with the traditional free electron  model~\cite{kittel}.

To prove that the EPO provides additional kinetic energy to thermally excited electrons in metals, we run a 100-ps equilibrium AIMD for Al at 329.40\,K and Li at 283.97\,K (both with a 2$\times$2$\times$2 conventional-cell and 32 total atoms). When $T>0\,\mathrm{K}$, the total energy of the free electron system can be written as~\cite{kittel,William}(SI Sec.~3)
\begin{eqnarray}
\nonumber E_{sys}=E_{0}+E_{T}=E_{0}+\frac{\pi^{2}}{4}\cdot N \frac{(k_{B}T)^{2}}{E_{F}^{0}},
\end{eqnarray}
where $E_{0}$ is the total energy of the free electron system at $0\,\mathrm{K}$, $E_{T}$ is the thermally excited energy of the free electron system obtained from the outside environment when $T>0\,\mathrm{K}$, $N$ is the total number of free electrons, and $E_{F}^{0}$ is the Fermi energy at $0\,\mathrm{K}$. We also calculate the energy provided by EPO using
\begin{eqnarray}
\nonumber E_{EPO}=2\cdot\overline{U}_{EPO}\cdot N\cdot e,
\end{eqnarray}  
where $\overline{U}_{EPO}$ is the average effective EPO amplitude. For Al, $E_{T}=2.6293\times 10^{-21}\,\mathrm{J}$ and $E_{EPO}=2.7674\times 10^{-21}\,\mathrm{J}$. For Li, $E_{T}=1.6049\times 10^{-21}\,\mathrm{J}$ and $E_{EPO}=1.5909\times 10^{-21}\,\mathrm{J}$. From these results, we have
\begin{eqnarray}
E_{T} \approx E_{EPO}.
\end{eqnarray}
We conclude that lattice vibrations cause EPO in metals, and simultaneously, EPO drives the collective vibration of free electrons. The energy of these collective vibrations provides additional kinetic energy to the thermally excited electrons. This is the core concept underlying this methodology.

Within this theoretical framework, higher temperatures strengthen the spatial EPO. To prove this relationship, we perform direct FFT of the relative displacement of EP $U_{\rm ion}$ and VEPO $\Delta U_{\rm ion}$. $U_{\rm ion}$ and $\Delta U_{\rm ion}$ were used to calculate $\SD{U}$ and $\SD{\Delta U}$ in Figs.~\ref{fig:2}($a~-~d$). $U_{\rm ion}$ describes the strength of the EPO in space, whereas $\Delta U_{\rm ion}$ reflects how fast the oscillation changes. Figs.~\ref{fig:2}(g,~h) show the frequency-dependent FFT amplitudes of $U_{\rm ion}$ and $\Delta U_{\rm ion}$, respectively. Clearly, the EPO is stronger and faster at higher temperatures.
 
In Fig.~\ref{fig:2}(i), we present the positive and negative integrations of the total $\Delta U_{\rm ion}$ in the same atom layer with simulation
time, which can be written as $\sum_{j=1}^{N_{al}}\sum_{t}\Delta U_{j}(t)$, where $j$ is the index of the atom in the layer and $N_{al}$ is the total number of atoms per layer. The four quantities in Fig.~\ref{fig:2}(i) show perfect linear behaviour over time. From the absolute values, we have:
\begin{equation}
\nonumber \sum_{j=1}^{N_{al}}\sum_{t}\vert\Delta U_{j}(t)_{hot}\vert>\sum_{j=1}^{N_{al}}\sum_{t}\vert\Delta U_{j}(t)_{cold}\vert ,
\end{equation}
which is consistent with the evidence shown in Figs.~\ref{fig:2}(g,~h). Notably, in the same temperature region, the positive and negative accumulations of $\sum_{j=1}^{N_{al}}\Delta U_{j}(t)$ are almost the same. In other words, $\sum_{j=1}^{N_{al}}\sum_{t}\Delta U_{j}(t)\simeq 0$, and thus, there is no net electric field gradient along the heat flux direction for a sufficiently long statistical time. This result confirms the physical picture illustrated in Fig.~\ref{fig:2}(f). Fig.~\ref{fig:2}(j) presents the distribution of the average effective amplitude of EPO in each atom
layer along the heat flux direction, $\overline{U}_{EPO}(l)$, where $l$ is the index of the atom layers. Moreover, the amplitude distribution of EPO explains how the thermal kinetic energy of thermally excited electrons is divided in space. We calculate $\overline{U}_{EPO}(l)$ using the RMS method~\cite{John}:
\begin{eqnarray}
\nonumber \overline{U}_{EPO}(l)=\frac{1}{N_{al}}\sum_{j=1}^{N_{al}}\sqrt{\frac{1}{n_{steps}}\sum_{t_{i}}^{n_{steps}}(U_{j}(t_{i})-\overline{U}_{j})^2},
\end{eqnarray} 
where $n_{steps}$ is the total number of simulation time steps, $U_{j}(t_{i})$ is the $U$ value of atom $j$ in a specific layer at time step $t_{i}$, and $\overline{U}_{j}$ is the average value of $U_{j}(t_{i})$. Then, we define the heat flux of electrons $\vec{J}_{el}$ according to the kinetic energy of thermally excited electrons between two adjacent atom layers. Because of the isotropy of the free electron model (SI Sec.~4), we take half of the difference of the thermal kinetic energy of thermally excited electrons between the two layers as
\begin{eqnarray}
\nonumber \vec{J}_{el}&=&-\frac{1}{2}\frac{n(e)\cdot e}{S\cdot t}\frac{\partial[2\cdot\overline{U}_{EPO}(l)\cdot n_{steps}]}{\partial N_{l}} \\
  &=&-\frac{n(e)\cdot e\cdot n_{steps}}{S\cdot t}\frac{\partial\overline{U}_{EPO}(l)}{\partial N_{l}},
\end{eqnarray}
where $S$ is the cross-sectional area, $t$ is the total simulation time, $n(e)$ is the number of free electrons per atom layer, and $\frac{\partial\overline{U}_{EPO}(l)}{\partial N_{l}}$ is the gradient of the average effective amplitude value of EPO by linear fitting of $\overline{U}_{EPO}(l)$ with the atom layer index number $N_{l}$ shown in Fig.~\ref{fig:2}(j). Here, a nonlinear phenomenon exists in the effective EPO amplitude distribution along the heat flux direction in some metals, such as Al, Be, and Mg. According to a case study of Be, we find that the nonlinear effect of $\overline{U}_{EPO}(l)$ can be reduced by increasing the system size (SI Sec.~4.2). Because of the nonlinear effect, when we calculate the $\vec{J}_{el}$ of Al, Be and Mg, we fit the linear part only. For Cu and Li, the $\overline{U}_{EPO}(l)$ distributions exhibit perfect linear behaviour along the heat current direction. Thus, we can calculate $\kappa_{el}$ based on Fourier's law:
\begin{eqnarray}
\vec{J}=-\kappa \nabla T.
\end{eqnarray}
Combining Eqs.~(5,~6), we obtain the expression for $\kappa_{el}$:
\begin{eqnarray}
\kappa_{el}=\frac{n(e)\cdot e \cdot n_{steps}}{\nabla T \cdot S\cdot t}\frac{\partial\overline{U}_{EPO}(l)}{\partial N_{l}},
\end{eqnarray}
where $\nabla T$ is obtained by linear fitting the temperature profile with the representative case shown in Fig.~\ref{fig:1}(d).

Within this framework, we studied the $\kappa_{el}$ of five metals (Li, Be, Mg, Al, and Cu) near room temperature. Additionally, by integrating the M\"uller-Plathe~\cite{Muller-Plathe} atomic kinetic energy flux, as shown in Fig.~\ref{fig:3}(b), we predict the lattice (phonon) thermal conductivities of the metals ($\kappa_{ph}$). Because of finite size effects, our NEAIMD results underestimate $\kappa_{ph}$, especially for Cu and Mg (SI Sec.~5). By summing $\kappa_{el}$ and $\kappa_{ph}$ from the NEAIMD simulations, we obtain the total thermal conductivities of the metals, as presented in Fig.~\ref{fig:3}(c). The results demonstrate that the thermal conductivities of metals slowly decrease with temperature near room temperature, which is consistent with traditional theory and experimental data. The error estimates in Fig.~\ref{fig:3}(c) are calculated from the expression for $\kappa_{el}$ and error propagation theory~\cite{Ku}. They mainly stem from the calculation of the gradient of $\overline{U}_{EPO}(l)$ and $\nabla T$. Here, we note that because the statistical temperature fluctuation $\overline{(\Delta T)^{2}}=k_{B}T^{2}/C_{v}$~\cite{Landau} of each atom layer is large (because of the small number of atoms per layer), the conventional error estimate of $\nabla T$ will be quite large. However, NEAIMD consistently yields a stable
temperature profile after a sufficiently long simulation time. Thus, we adopt the error in the linear fitting for $\nabla T$. We also note that the
aforementioned nonlinear phenomenon of the gradient of $\overline{U}_{EPO}(l)$ can also lead to large error bars. The details of the error-bar analysis can be found in SI Sec.~6.

In Fig.~\ref{fig:3}(d), our results show that $\kappa_{el}$ indeed dominates the thermal transport process in metals. To compare our results with those of the traditional BTE method, we also utilize the BoltzTraP software~\cite{Madsen} (based on electron energy band theory) to calculate $\frac{\kappa_{el}}{\tau_{el}}$. In obtaining $\kappa_{el}$, we use the constant relaxation time approximation $\tau_{el}=1\times 10^{-14}~\mathrm{s}$~\cite{Madsen,Chen}. To avoid finite size effects in the calculation of $\kappa_{ph}$, we also evaluate $\kappa_{ph}$ from the BTE method with interatomic force constants obtained from ab initio calculations~\cite{broido,chengang}, as implemented in the ShengBTE package~\cite{Wu}. Then, we obtain the total thermal conductivities of the metals via the BTE method by summing $\kappa_{el}$ from BoltzTraP and $\kappa_{ph}$ from ShengBTE. Our NEAIMD method, the traditional BTE method and experimental data are compared in Fig.~\ref{fig:3}(e). The results demonstrate that our method is superior to the traditional BTE method in predicting the electronic thermal conductivities of metals, especially for Be and Cu at room temperature. Moreover, we observe an interesting phenomenon when calculating the spectral density of VEPO ($\SD{\Delta U}$) in Fig.~\ref{fig:2}(b,~d). We perform exponential decay fitting of the autocorrelation function of the VEPO using the formula $y=A\mathrm{exp}(-t/\tau_{\mathrm{exp}})$. Surprisingly, the exponential autocorrelation time of VEPO $\tau_{\mathrm{exp}}$ at room temperature is on the same approximate order of magnitude as the theoretical collision time of the free electrons~\cite{Madsen,Chen,Campillo}. The results for Cu and Al are shown in Fig.~\ref{fig:3}(a). We also examine other metals (Be, Li, and Mg) and obtain similar results (SI Sec.~8). Therefore, we anticipate that some physical mechanisms must drive this phenomenon, i.e., it is not a coincidence.

In summary, we have developed a new methodology based on the concept of EPO to predict the electronic thermal conductivities of metals via direct NEAIMD simulation. Without explicitly addressing any complicated scattering processes of free electrons, our NEAIMD-EPO method provides better predictions of the electronic thermal conductivities of pure metals than the traditional BTE method near room temperature. We expect that this methodology will be helpful and useful for understanding and studying the heat-transfer problems of metal systems in the future. Further extension to cope with some presently challenging problems in materials, such as electron-phonon coupling, is also foreseen.

\section*{Acknowledgments}
M.H. gratefully acknowledges Prof.\ David G. Cahill (University of Illinois at Urbana-Champaign) for valuable comments on the manuscript. S.Y.Y. gratefully thanks Dr.\ Yang Han (RWTH Aachen University) and Dr.\ Tao Ouyang (Xiangtan University) for their helpful and fruitful discussions, and Dr.\ Shi-Ju Ran (ICFO, Spain) for his help in plotting Figs.~\ref{fig:1}(b,~e). X.Z. greatly acknowledges Dr. Zhiwei Cui (Northwestern University) for providing the input script and MEAM potential files for the EMD simulations of Li. The authors gratefully acknowledge the computing time granted by the John von Neumann Institute for Computing (NIC) and provided on the supercomputer JURECA at J\"ulich Supercomputing Centre (JSC) (Project ID: JHPC25). The classical NEMD and EMD simulations were performed with computing resources granted by the J\"ulich Aachen Research Alliance-High Performance Computing (JARA-HPC) from RWTH Aachen University under project No. jara0135. S.S. was supported by NERC grant number NE/K006290/1.

\section*{Author Contributions}
M.H. and S.Y.Y conceived the project. M.H. supervised the project and E.D.N. co-supervised the project. S.Y.Y. developed the methodology and theoretical formula. S.Y.Y designed and executed all the AIMD simulations and calculations of the electronic thermal transport properties. G.Q., X.Z. and S.Y.Y. post-processed and analyzed the simulation data. X.Z. and S.Y.Y. performed the frequency domain analysis. S.S. provided the original code of NEAIMD. X.Z. and S.Y.Y improved the NEAIMD code. S.Y.Y. calculated the thermal conductivity of metals with the BTE method. S.Y.Y, X.Z., E.D.N., S.S and M.H. co-wrote the manuscript. All authors participated in the discussions and reviewed and revised the manuscript.

\section*{Additional Information}
Competing financial interests: The authors declare no competing financial interests.

\newpage


\section*{Supplementary material}

## 1. NEAIMD simulation step

All AIMD simulations are performed using the DFT method implemented in VASP [1, 2]. The Perdew-Burke-Ernzerhof parameterization of the generalized gradient approximation (GGA) is used for the exchange-correlated functional [3], and the projector-augmented wave method is applied to model the core electrons [4, 5]. For the energy cut-off, we use the default value in the pseudopotential file. We first relax the system with the *NpT*-ensemble (constant number of particles, pressure, and temperature) at room temperature to obtain the lattice constants of the metals at a finite temperature. We then apply the resulting lattice constants to construct the initial structures for subsequent NEAIMD simulations.

The NEAIMD simulations are performed using a modified version of the VASP code [6] with the *NVE* ensemble (constant volume and no thermostat). We apply a fixed boundary condition along the direction of the heat flux and periodic boundary conditions in the two lateral directions (perpendicular to the direction of the heat flux) of the simulation model (see Fig. 1(c) in the main article). To avoid self-interaction between periodic images of the simulation cell, in the direction of the heat flux, we add a vacuum layer with a thickness exceeding 5 Å on the external sides of the fixed atom layers. The total distance between the periodic images of the simulation model exceeds 10 Å in real space. The layers next to the fixed layers are the heat-source and heat-sink. A constant atomic heat flux is imposed by applying the Müller-Plathe algorithm [7]. The coldest atom in the hot region and the hottest atom in the cold region are selected, and their kinetic energies (atomic velocities) are exchanged every 50 fs with a 1-fs time step. This operation induces a steady heat-energy-flux in the system and a corresponding temperature gradient ($\nabla T$) after running for a sufficiently long time. The energy-exchange time interval is used to control the temperature gradient's magnitude. The linear portion of the temperature gradient lies between the heat baths. By linear fitting the statistically averaged temperatures of each atom layer, we obtain $\nabla T$, which is used to calculate the final electronic thermal conductivity ($\kappa_{el}$) and phonon thermal conductivity ($\kappa_{ph}$).

For Li, Al and Cu, the size of simulation model is $2 \times 2 \times 8$ conventional cells (16 atom layers along the heat flux direction). For Be and Mg, the size of the simulation model is $2 \times 4 \times 8$ primitive cells (16 atom layers along the heat flux direction). To study size effects, we examine models with a length of 24 atom layers for each metal, i.e., $2 \times 2 \times 12$ conventional cells for Li, Al and Cu and $2 \times 4 \times 12$ primitive cells for Be and Mg. For Al, we also investigate the size effect of the cross-section using a model with $4 \times 4 \times 8$ conventional cells. All the simulation temperatures are approximately 300 K. The basic information for all NEAIMD simulations is presented in Table 1. Additionally, a short movie of the NEAIMD of Cu is given to show the real simulation process.

Table 1: Details of NEAIMD simulation of metals

| System | System length (Å) | Cross-sectional area ($S$, Å$^2$) | Atom number | Total simulation time (t, ps) | Average temperature (T, K) | ID of simulation cases |
|---|---|---|---|---|---|---|
| Li | 38.9160 | 74.7983 | 128 | 100 | 288.20 | Li-1 |
|  |  |  |  |  | 293.08 | Li-2 |
|  |  |  |  |  | 303.64 | Li-3 |
|  |  |  |  |  | 305.35 | Li-4 |
|  |  |  |  |  | 313.87 | Li-5 |
|  |  |  |  |  | 316.28 | Li-6 |
| Be | 45.8631 | 36.0108 | 192 | 70.503 | 278.02 | Be-1 |
|  | 31.7520 |  | 128 | 92.206 | 285.00 | Be-2 |
|  |  |  |  | 95.996 | 296.17 | Be-3 |
|  |  |  |  | 95.320 | 311.22 | Be-4 |
|  |  |  |  | 95.907 | 313.46 | Be-5 |
| Mg | 47.2410 | 70.2266 | 128 | 64.906 | 290.95 | Mg-1 |
|  |  |  |  | 65.178 | 297.74 | Mg-2 |
|  | 68.2388 |  | 196 | 53.513 | 302.19 | Mg-3 |
|  |  |  |  | 64.520 | 307.87 | Mg-4 |
|  | 47.2410 |  | 128 | 63.368 | 311.41 | Mg-5 |
| Al | 52.5067 | 65.2525 | 192 | 51.958 | 284.48 | Al-1 |
|  |  |  |  | 76.497 | 286.14 | Al-2 |
|  |  |  |  | 75.340 | 299.46 | Al-3 |
|  | 36.3510 |  | 128 | 77.251 | 306.84 | Al-4 |
|  |  |  |  | 71.534 | 310.39 | Al-5 |
|  |  |  |  | 73.971 | 318.31 | Al-6 |
| Cu | 47.0096 | 52.3134 | 196 | 13.424 | 285.31 | Cu-1 |
|  | 32.5440 |  | 128 | 22.280 | 293.86 | Cu-2 |
|  |  |  |  | 22.787 | 298.49 | Cu-3 |
|  |  |  |  | 22.719 | 310.91 | Cu-4 |

## 2. PSD and the autocorrelation function

The power spectrum $S_x(f)$ of a time series $x(t)$ describes the distribution of the frequency components composing that signal. According to Fourier analysis, any physical signal can be decomposed into a number of discrete frequencies or a continuous spectrum of frequencies. The statistical average of a certain signal or signal type (including noise), analysed regarding its frequency content, is called its spectrum [8, 9].

The autocorrelation function of a real stationary signal $x(t)$ is defined as

$$R_x(\tau) = E[x(t)x(t+\tau)], \qquad (1)$$

where $E[\cdot]$ is the expected value operator. The Fourier transform of $R_x(\tau)$ is called the PSD $S_x(f)$ [8, 9]

$$S_x(f) = \int_{-\infty}^{\infty} R_x(\tau)e^{-2\pi i f\tau}d\tau. \tag{2}$$

Given any two frequencies $f_1$ and $f_2$, the quantity

$$\int_{f_1}^{f_2} S_x(f)df \tag{3}$$

represents the portion of the average signal power contained in the signal frequencies from $f_1$ to $f_2$, where $S_x$ is a spectral density.

Here, to demonstrate the relationship between lattice vibration (i.e., thermal motion of the ion cores) and EPO at the ion cores, we consider the following four physical signals:

1) The displacement of the ion cores $D_{ion}(t)$. We extract the position of each ion core from the AIMD run at each time step from the VASP output file. Here, we are particularly interested in the $\vec{z}$ component, i.e., that along the direction of the heat flux. By applying the PSD technique, we obtain the spectral density of atomic displacement ($S_D$), which gives information regarding atomic vibrations in real space.
2) The velocity of the ion cores $V_{ion}(t)$. We modified the VASP code to output the instantaneous velocity of each ion core at every AIMD time step. Analogous to $D_{ion}(t)$, here, we take the $\vec{z}$ component for $V_{ion}(t)$. By applying the PSD technique we obtain the spectral density of atomic velocity ($S_V$). $S_V$ is conventionally known as the vibrational density of states (VDOS) and provides information regarding the oscillation velocity of the atoms.
3) The EP displacement at the ion cores $U_{ion}(t)$. We extract the EP value at each ion core at every AIMD time step from the VASP output file. By applying the PSD technique, we obtain the spectral density of EP displacement ($S_U$). Based on its definition, $S_U$ reflects the behaviour of EPO at the ion core in real space.
4) The VEPO at the ion cores $\Delta U_{ion}(t)$. We define $\Delta U_{ion}(t)$ as the difference between $U_{ion}(t)$ at time step $t + 1$ and $t$. By applying the PSD technique, we obtain the spectral density of VEPO ($S_{\Delta U}$). Similar to $S_V$ (VDOS), $S_{\Delta U}$ provides information about the rate of the EP value change at the ion core.

The EP expression used in VASP is defined as

$$U = \int U(r) \cdot \rho_{test} \cdot (|r - R|)d^3r, \tag{4}$$

where the test charge $\rho_{test}$ is norm 1. From this formula, we can see that the EP $U$ is a function of ion position $R$. Thus, the EPO at the ion core is induced by the lattice vibrations, i.e., the thermal motion of the ion cores. Therefore, in principle, $S_D$ and $S_U$ should have some correlation, similar to $S_V$ and $S_{\Delta U}$.

To prove our conjecture, for a selected model of Al, we performed an additional 10-ps equilibrium AIMD simulation with the *NVE* ensemble. The results are shown in Fig. 2(a, b) in the main text. Clearly, the distributions of $S_D$ and $S_U$ and of $S_V$ and $S_{\Delta U}$ are very similar, and the locations of the peaks, with respect to frequency, are in exact agreement. These results strongly support our conjecture. Similarly, the same quantities were determined via a 100-ps NEAIMD

simulation of Al (Fig. 2(c, d)), which revealed a strong correlation between the lattice vibration and the EPO at the ion cores. However, we also note that for some specific frequencies (3.5~5.0 THz), the peaks of the spectral density from the lattice vibration and EPO in Figs. 2(c, d) are not exactly the same. We suspect that this phenomenon may be attributable to the heat flux applied in NEAIMD. Nevertheless, comparing the PSD results confirms that the relationship between lattice vibration and EPO holds.

## 3. Evidence for the EPO from lattice vibrations providing the thermal kinetic energy to thermally excited electrons

We first give a detailed derivation of the total energy expression for a free electron system at T > 0 K.

From the Sommerfeld expansion [10], if the function Q(E) is continuously differentiable on $(-\infty, +\infty)$, $Q(0) = 0$, and $\lim_{E \to \infty} e^{-\alpha E} Q(E) = 0$ (here, α is a positive and constant number) when $k_B T \ll E_F$, then

$$I = \int_0^\infty f(E) Q'(E) dE \approx Q(E_F) + \frac{\pi^2}{6}(k_B T)^2 Q''(E_F), \tag{5}$$

where $f(E)$ is the Fermi-Dirac distribution function, $E_F$ is the Fermi energy, $k_B$ is the Boltzmann constant and $T$ is the system temperature.

When T > 0 K, the total energy of the free electron system $U_{sys}$ can be written as

$$U_{sys} = \int_0^\infty E f(E) N(E) dE$$

$$= \int_0^{E_F} E N(E) dE + \frac{\pi^2}{6}(k_B T)^2 \frac{d}{dE}[E N(E)]_{E_F}$$

$$= \int_0^{E_F^0} E N(E) dE + \int_{E_F^0}^{E_F} E N(E) dE + \frac{\pi^2}{6}(k_B T)^2 \cdot \frac{3}{2}(\frac{1}{2\pi^2}(\frac{2m_e}{\hbar^2})^{\frac{3}{2}} \cdot \sqrt{E})$$

$$\approx E_0 + E_F^0 N(E_F^0)(E_F - E_F^0) + \frac{\pi^2}{4}(k_B T)^2 N(E_F^0).$$

$$\because E_F = E_F^0 \left[1 - \frac{\pi^2}{12}\left(\frac{k_B T}{E_F^0}\right)^2\right] \text{ and } N(E_F^0) = \frac{3N}{2E_F^0}$$

$$\therefore U_{sys} = \int_0^{E_F^0} E N(E) dE + N(E_F^0)\left[-\frac{\pi^2}{12}(k_B T)^2\right] + \frac{\pi^2}{4}(k_B T)^2 N(E_F^0)$$

$$= \int_0^{E_F^0} E N(E) dE + \frac{\pi^2}{6}(k_B T)^2 N(E_F^0)$$

$$= E_0 + \frac{\pi^2}{4} \cdot N \frac{(k_B T)^2}{E_F^0}. \tag{6}$$

The first term ($E_0$) on the right-hand side of Equation (6) is the total energy of the free electron system at 0 K, whereas the second term is the thermally excited energy ($E_T$) of the system obtained from the outside environment when T > 0 K. Here, $N$ is the total number of free electrons, and $E_F^0$ is the Fermi energy at 0 K. Because the Fermi energy changes very little with temperature, here, we take $E_F$ at room temperature as $E_F^0$ [11].

In Fig. 2(f) of the main text, we demonstrate that the local vibrational EP field, which originates from the EPO, causes the collective vibration of free electrons in the momentum space and provides additional thermal kinetic energy to the thermally excited electrons near the Fermi surface. To confirm this, we perform two additional equilibrium AIMD simulations: Al (100 ps at 329.40 K) and Li (100 ps at 283.97 K). Both use a supercell of 2 × 2 × 2 conventional cells (32 atoms in total) and periodic boundary conditions in all three dimensions.

We define the total energy provided by EPO as

$$E_{EPO} = \sum_i 2 \cdot \bar{U}_i \cdot n_i(e), \tag{7}$$

where $\bar{U}_i$ is the average effective amplitude of EPO at the ion core $i$. Because the simulations here are equilibrium AIMD, the $\bar{U}_i$ of different atoms can be taken as having the same value (over a long time average, the temperature of each ion core can be assumed to be the same). $n_i(e)$ is the number of free electrons per atom (for Al, $n_i(e)=3$; for Li, $n_i(e) = 1$). Then,

$$E_{EPO} = \sum_i 2 \cdot \bar{U}_i \, n_i(e) = 2 \cdot \bar{U}_{EPO} \sum_i n_i(e) = 2 \cdot \bar{U}_{EPO} \cdot N, \tag{8}$$

where $N$ is the total number of free electrons in the system and $\bar{U}_{EPO}$ is the average effective amplitude of EPO from equilibrium AIMD

$$\bar{U}_{EPO} = \frac{1}{N_a} \sum_{j=1}^{N_a} \sqrt{\frac{1}{n_{steps}} \sum_{t_i}^{n_{steps}} (U_j(t_i) - \bar{U}_j)^2}, \tag{9}$$

where $N_a$ is the total number of atoms in the simulation cell, $n_{steps}$ is the total simulation steps, $U_j(t_i)$ is the EP value of atom $j$ in the cell at time step $t_i$, and $\bar{U}_j$ is the average EP value of atom $j$. $\bar{U}_{EPO}$ is calculated using the RMS method [12]. $E_T$ and $E_{EPO}$ for Al and Li are compared in Table 2.

Table 2: Comparison of $E_T$ and $E_{EPO}$ for Al and Li from equilibrium AIMD simulations near room temperature

| System | Temperature (K) | Total number of free electrons | Fermi energy (eV) | Average effective amplitude of EPO ($10^{-5}$V) | Thermally excited energy $U_T$ ($10^{-21}$ J) | Total energy provided by EPO $E_{EPO}$ ($10^{-21}$ J) |
|---|---|---|---|---|---|---|
| Li | 283.97 | 32 | 4.72 | 15.5152 | 1.6049 | 1.5909 |
| Al | 329.40 | 96 | 11.63 | 8.9964 | 2.6293 | 2.7674 |

From the results in Table 2, the following relationship is clear

$$E_T \approx E_{EPO}. \tag{10}$$

This result supports our conjecture that EPO causes the collective vibration of free electrons in momentum space and provides the thermal kinetic energy of the thermally excited electrons.

## 4. Heat flux via free electrons $\vec{J}_{el}$
### 4.1 Results of the EPO in space

As shown in Figs. 2(g, h) of the main text, higher temperatures can increase the strength and speed of spatial EPO. Figs. 2(i, j) in the main text confirm this. From Fig. 2(i), we also have

$$\sum_{j=1}^{N_{al}} \sum_t \Delta U_j(t) \cong 0. \tag{11}$$

Therefore, no net local electric field occurs over the simulation. In other words, no net electric current exists in metals during the heat-transfer process, in good agreement with our common sense arguments. We also note that both the negative and positive parts of $\sum_{j=1}^{N_{al}} \sum_t \Delta U_j(t)$ exhibit perfect linear behaviours with time, as shown in Fig. 2(i). We apply the RMS method to calculate the average effective amplitude of EPO ($\bar{U}_{EPO}$)

$$\bar{U}_{EPO}(l) = \frac{1}{N_{al}} \sum_{j=1}^{N_{al}} \sqrt{\frac{1}{n_{steps}} \sum_{i=1}^{n_{steps}} (U_j(t_i) - \bar{U}_j)^2}, \tag{12}$$

where $l$ is the index of the atom layers, $N_{al}$ is the total number of atoms per layer, $n_{steps}$ is the total number of simulation steps, $U_j(t_i)$ is the EP displacement of ion core $j$ at $i$ fs, and $\bar{U}_j$ is the average value of $U_j(t_i)$ for atom $j$ in layer $l$. $\bar{U}_{EPO}(l)$ represents the intensity of the local EPO. Thus, the energy provided by EPO in each layer $E_{EPO}(l)$ can be written as

$$E_{EPO}(l) = 2 \cdot \bar{U}_{EPO}(l) \cdot n(e) \cdot e \tag{13}$$

### 4.2 The distribution of thermally excited electrons' thermal kinetic energy along the heat flux direction and the non-linear effect analysis

We have proven that the energy from EPO provides the kinetic energy of thermally excited electrons. Therefore, we can treat $E_{EPO}$ as the thermal kinetic energy of thermally excited electrons. Combining Eqs. (12) and (13), we present the distribution of the effective amplitude of EPO in space rather than that of the thermal kinetic energy of thermally excited electrons, in Fig. S1.

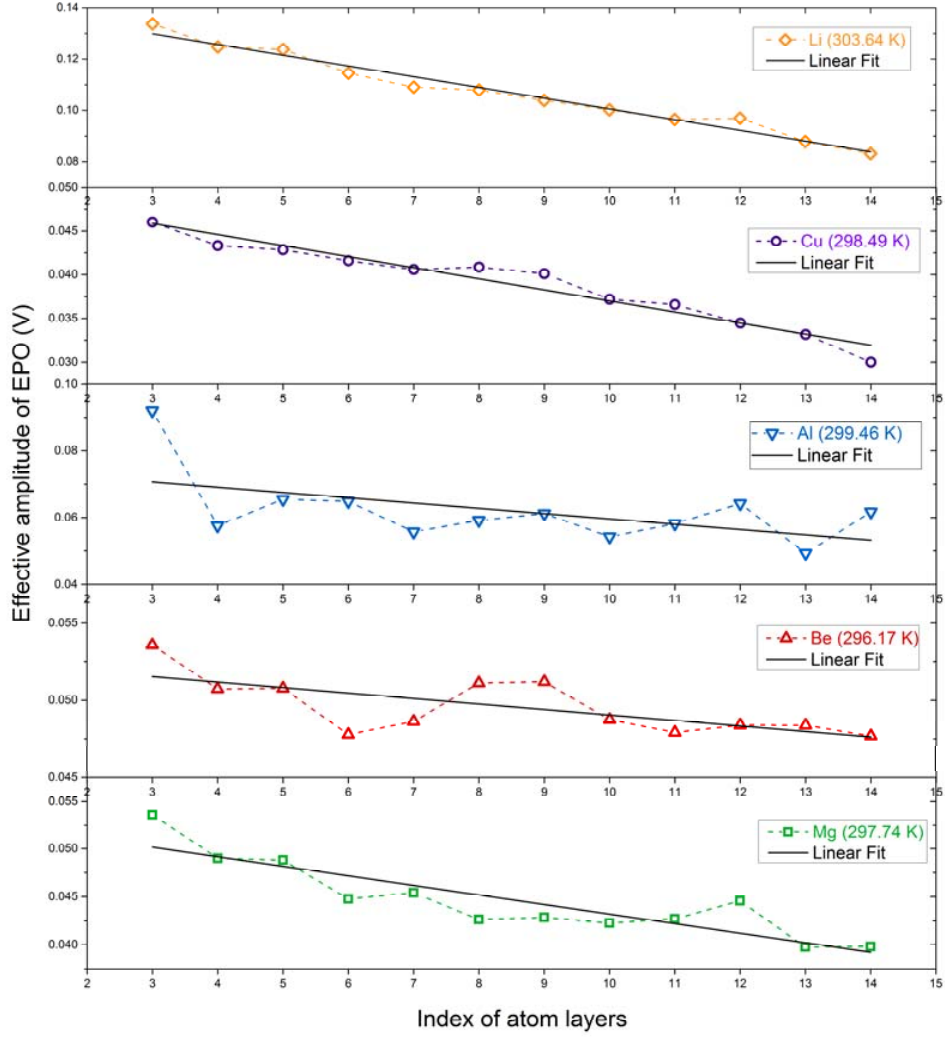

Fig. S1: Effective amplitudes of the EPO along the direction of the heat flux in different metals (all systems are approximately 300 K). Via linear fitting, we can obtain $\frac{\partial \bar{U}_{EPO}(l)}{\partial N_l}$, which is used to calculate electronic heat flux.

Fig. S1 shows that in metals, higher temperatures increase the strength of EPO. We also note that a non-linear phenomenon of $\bar{U}_{EPO}(l)$ occurs in some metals, such as Al, Be. This phenomenon will cause significant errors in the final electronic thermal conductivity ($\kappa_{el}$).

To elucidate the reason for this non-linear phenomenon, we plot the temperature profiles of different metals in Fig. S2. The degrees of the non-linear temperature distributions of Be and Al are larger than those of Li, Cu, and Mg at approximately 300 K. The non-linear temperature distribution may be responsible for the non-linear distribution of $\bar{U}_{EPO}(l)$.

8

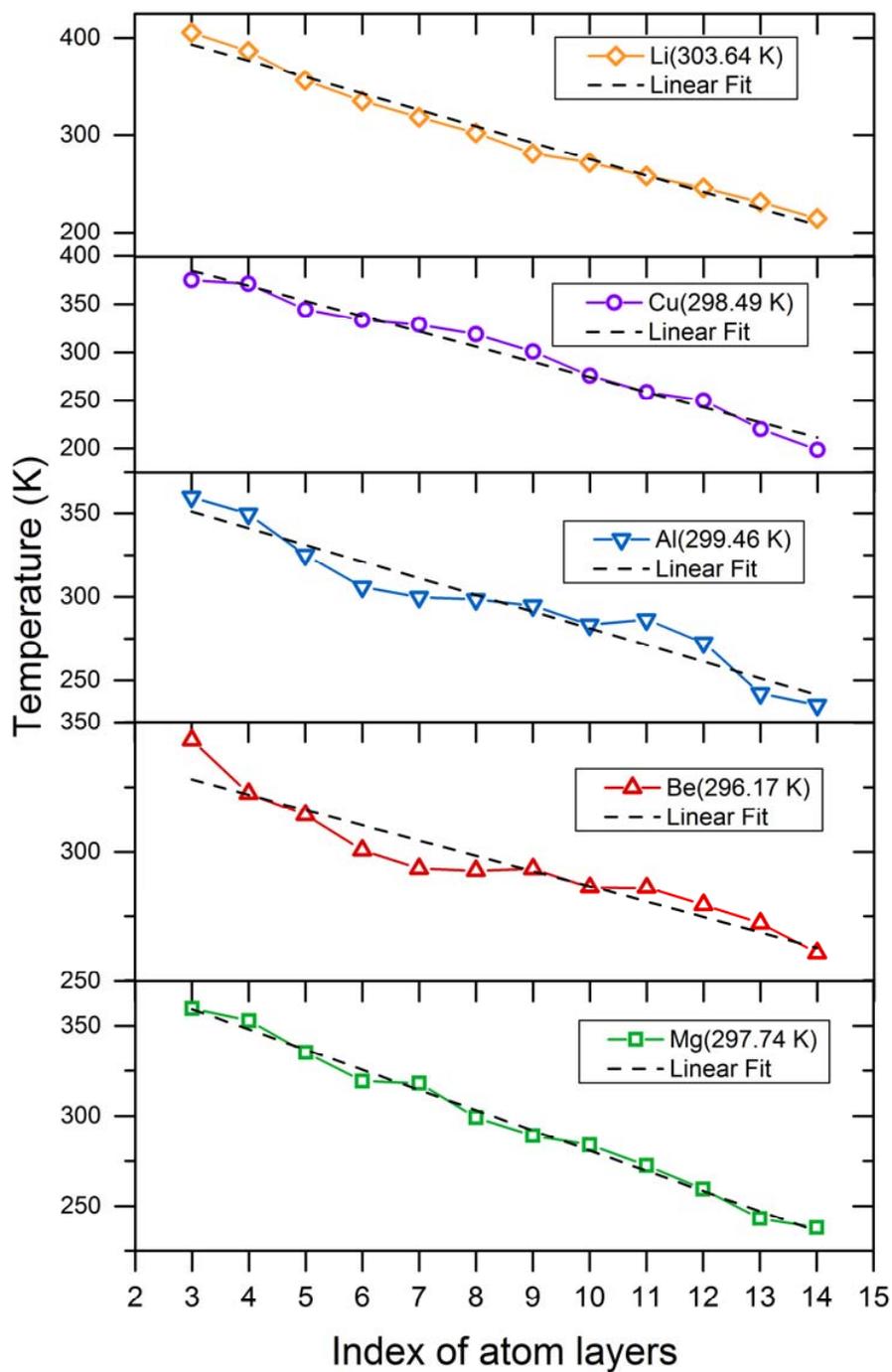

Fig. S2: Temperature profiles from NEAIMD simulations of different metals (all systems are at approximately 300 K). The dashed line is the linear fit of the temperature profile, i.e., the temperature gradient ($\nabla T$).

We also attempt to reduce the non-linear effect by increasing the model size. Specifically, we vary the sizes of the Be and Al models. Fig. S3 presents the non-linear effects of different Be lengths at approximately 300 K. Fig. S4 shows the non-linear effects of different cross-sections and lengths of Al at approximately 300 K. Based on these results, the simulation size indeed affects the non-linear EPO phenomenon in metals, and the non-linear effects of EPO along the heat flux direction can be reduced by increasing the simulation size. However, we could not completely eliminate the non-linear effects by increasing the simulation size. The mechanism underlying this non-linear phenomenon warrants further study.

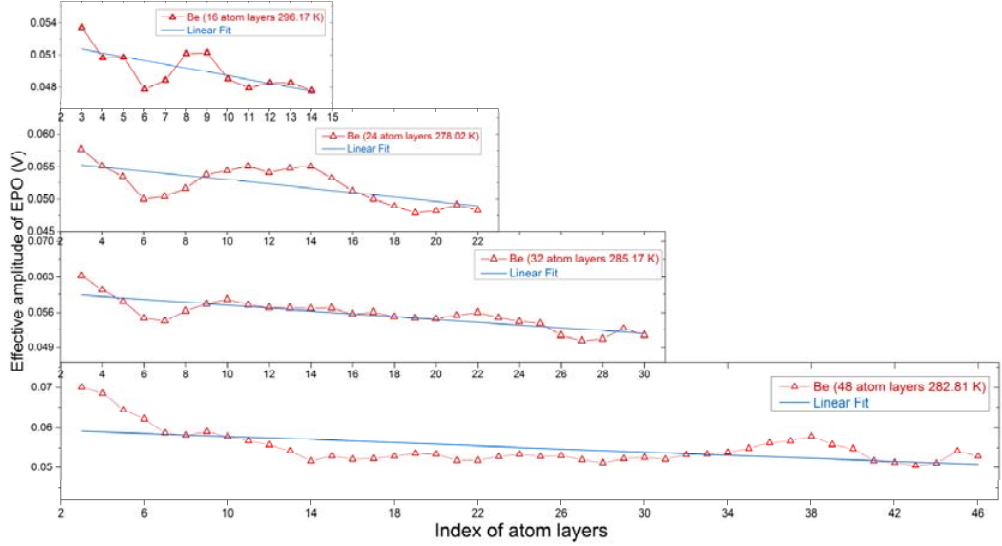

Fig. S3: Distributions of the effective EPO amplitude for different lengths of Be (all systems are at approximately 300 K).

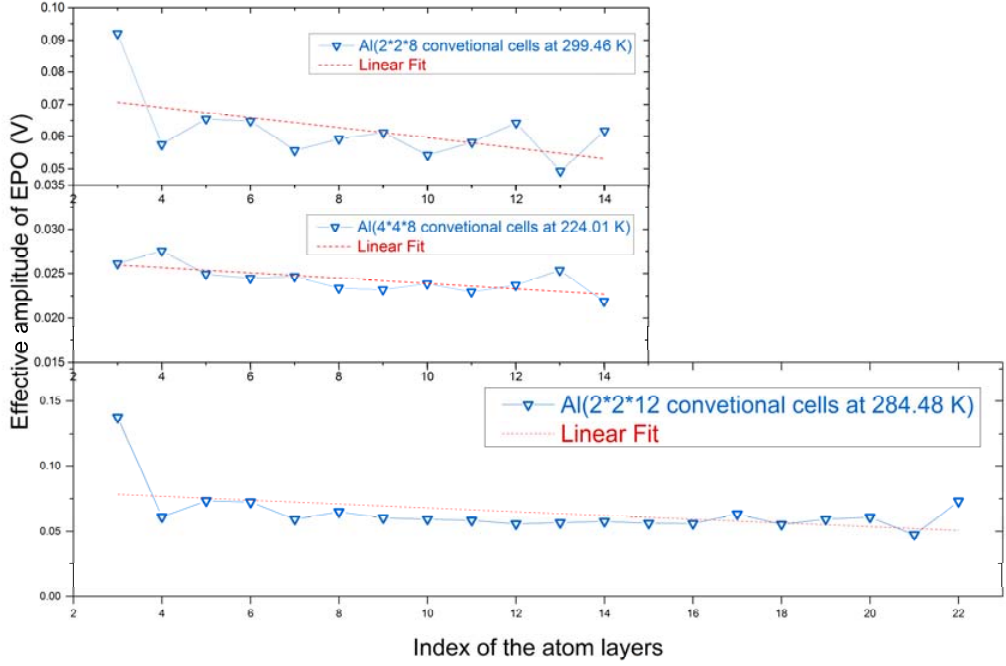

Fig. S4: Distributions of the effective EPO amplitude for different cross-sections and lengths of Al (all systems are at approximately 300 K).

### 4.3 Definition of $\vec{J}_{el}$

First, we calculate the average effective EPO amplitude $\bar{U}_{EPO}(l)$ and effective EPO energy $E_{EPO}(l)$. Then, we can obtain the total effective energy provided by EPO during simulation time $t$ as

$$E_{EPO}(l,t) = 2 \cdot \bar{U}_{EPO}(l) \cdot n(e) \cdot e \cdot n_{steps}, \tag{14}$$

where $n_{steps}$ is the total number of time steps during the simulation time $t$. When the system reaches a quasi-equilibrium state, we can infer that the thermal energy of thermally excited electrons is exchanged between two adjacent atom layers. As illustrated in Fig. S5, we take half of the difference of the thermal kinetic energy exchange between the two layers as $\vec{J}_{el}$ (because of the isotropy of the free electron model)

$$\begin{aligned}\vec{J}_{el} &= -\frac{1}{2}\frac{n(e) \cdot e}{S \cdot t}\frac{\partial\left[2 \cdot \bar{U}_{EPO}(l) \cdot n_{steps}\right]}{\partial N_l} \\ &= -\frac{n(e) \cdot e \cdot n_{steps}}{S \cdot t}\frac{\partial \bar{U}_{EPO}(l)}{\partial N_l}\end{aligned}, \tag{15}$$

11

where $S$ is the cross-sectional area, $n(e)$ is the number of free electrons per atom layer, $e$ is the unit charge of a single electron, and $\frac{\partial \bar{U}_{EPO}(l)}{\partial N_l}$ is the gradient of the average effective EPO amplitude value by linear fitting of $\bar{U}_{EPO}(l)$ with respect to the index number of atom layers ($N_l$), as shown in Fig. S1.

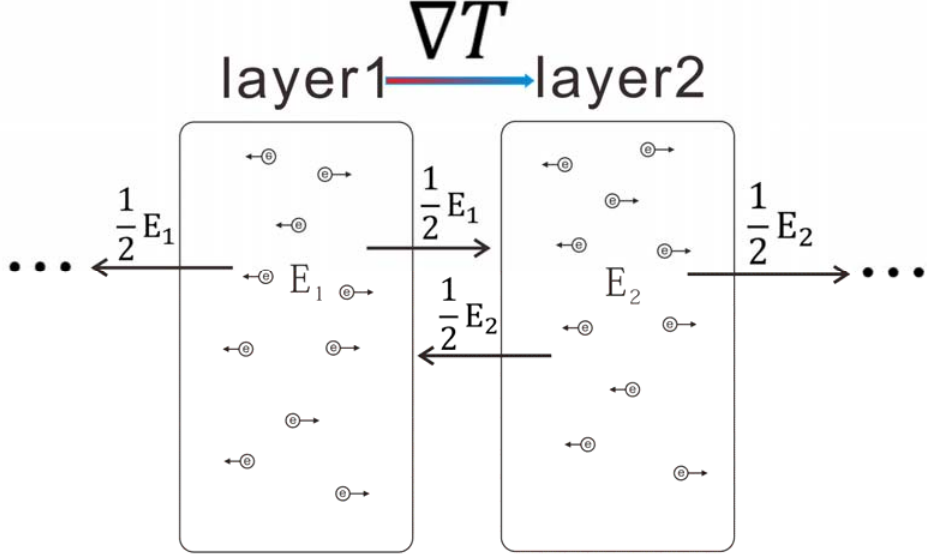

Fig. S5: Schematic of the exchange of thermal excited electrons' thermal energy between two adjacent atom layers. $E_1$ and $E_2$ are the thermal energies carried by thermally excited electrons in layer1 and layer2, respectively.

## 5. Calculation of thermal conductivity κ of metals from NEAIMD
### 5.1 Electronic thermal conductivity ($\kappa_{el}$)

From Fourier's Law of heat conduction, the electronic thermal conductivity ($\kappa_{el}$) can be written as

$$\kappa_{el} = -J_{el}/\nabla T. \tag{16}$$

Combining Equation (16) with Equation (15), we have

$$\kappa_{el} = \frac{n(e)\cdot e\cdot n_{steps}}{S\cdot t\cdot \nabla T}\frac{\partial \bar{U}_{EPO}(l)}{\partial N_l}. \tag{17}$$

Based on Eq. (17), we calculate the $\kappa_{el}$ of five metals: Li, Be, Mg, Al, and Cu around room temperature. We run multiple NEAIMD simulations for each system with different $T$ to examine the temperature dependent thermal conductivity of metals. All the simulation results are reported in Table 3.

## 5.2 Phonon thermal conductivity ($\kappa_{ph}$)**5.2 Phonon thermal conductivity ($\kappa_{ph}$)**

As we employ the Müller-Plathe algorithm to establish a stable temperature gradient along the heat transfer direction, we can also obtain the atomic kinetic energy flux $J_{ph}$. We calculated the phonon thermal conductivity ($\kappa_{ph}$) simultaneously from Fourier's Law:

$$\kappa_{ph} = -J_{ph}/\nabla T. \tag{18}$$

The results are also reported in Table 3. By summing $\kappa_{el}$ and $\kappa_{ph}$, we obtain the total $\kappa$ of metals from parameter free NEAIMD simulations.

Table 3: Electronic thermal conductivity $\kappa_{el}$ and phonon thermal conductivity $\kappa_{ph}$ from NEAIMD.

| System | Case ID | Average temp. (T, K) | Temperature gradient $\nabla T$ (K/Å) | Electronic thermal conductivity $\kappa_{el}$ (W/mK) | Phonon thermal conductivity $\kappa_{ph}$ (W/mK) | Total thermal conductivity $\kappa$ (W/mK) |
|---|---|---|---|---|---|---|
| Li | Li-1 | 288.20 | -7.2117 | 82.1294 | 2.7502 | 84.8796 |
|    | Li-2 | 293.08 | -7.6875 | 81.0457 | 2.6219 | 83.6676 |
|    | Li-3 | 303.64 | -7.7978 | 79.9341 | 2.6386 | 82.5727 |
|    | Li-4 | 305.35 | -7.9970 | 78.4404 | 2.5930 | 81.0334 |
|    | Li-5 | 313.87 | -8.1606 | 77.3224 | 2.6050 | 79.9274 |
|    | Li-6 | 316.28 | -8.2397 | 76.7194 | 2.5927 | 79.3121 |
| Be | Be-1 | 278.02 | -2.2042 | 220.9939 | 25.0928 | 246.0867 |
|    | Be-2 | 285.00 | -2.9479 | 216.3805 | 21.0501 | 237.4306 |
|    | Be-3 | 296.17 | -3.3623 | 196.5404 | 18.8916 | 215.4320 |
|    | Be-4 | 311.22 | -3.8448 | 174.4980 | 16.9305 | 191.4285 |
|    | Be-5 | 313.46 | -3.6524 | 164.2614 | 18.5753 | 182.8367 |
| Mg | Mg-1 | 290.95 | -4.8223 | 166.4749 | 2.1260 | 168.6009 |
|    | Mg-2 | 297.74 | -4.2645 | 148.2052 | 2.4020 | 150.6072 |
|    | Mg-3 | 302.19 | -3.6466 | 140.7474 | 2.3877 | 143.1351 |
|    | Mg-4 | 307.87 | -5.8340 | 140.4684 | 1.7672 | 142.2356 |
|    | Mg-5 | 311.41 | -5.1873 | 142.9120 | 2.0244 | 144.9364 |
| Al | Al-1 | 284.48 | -3.0485 | 232.7718 | 5.7530 | 238.5248 |
|    | Al-2 | 286.14 | -4.9667 | 231.2791 | 4.5353 | 235.8144 |
|    | Al-3 | 299.46 | -4.9448 | 223.9373 | 4.6760 | 228.6133 |
|    | Al-4 | 306.84 | -5.2123 | 223.6268 | 4.6600 | 228.2868 |
|    | Al-5 | 310.39 | -5.3456 | 220.0350 | 4.3882 | 224.4232 |
|    | Al-6 | 318.31 | -5.5119 | 209.5108 | 4.3312 | 213.8420 |
| Cu | Cu-1 | 285.31 | -7.2400 | 438.9252 | 2.0393 | 440.9645 |
|    | Cu-2 | 293.86 | -9.6536 | 399.0878 | 1.8816 | 400.9694 |
|    | Cu-3 | 298.49 | -8.7284 | 391.4515 | 2.1829 | 393.6344 |
|    | Cu-4 | 310.91 | -9.9209 | 374.4141 | 1.9523 | 376.3664 |

**5.2 Phonon thermal conductivity ($\kappa_{ph}$)**

As we employ the Müller-Plathe algorithm to establish a stable temperature gradient along the heat transfer direction, we can also obtain the atomic kinetic energy flux $J_{ph}$. We calculated the phonon thermal conductivity ($\kappa_{ph}$) simultaneously from Fourier's Law:

$$\kappa_{ph} = -J_{ph}/\nabla T. \tag{18}$$

The results are also reported in Table 3. By summing $\kappa_{el}$ and $\kappa_{ph}$, we obtain the total $\kappa$ of metals from parameter free NEAIMD simulations.

Table 3: Electronic thermal conductivity $\kappa_{el}$ and phonon thermal conductivity $\kappa_{ph}$ from NEAIMD.

| System | Case ID | Average temp. (T, K) | Temperature gradient $\nabla T$ (K/Å) | Electronic thermal conductivity $\kappa_{el}$ (W/mK) | Phonon thermal conductivity $\kappa_{ph}$ (W/mK) | Total thermal conductivity $\kappa$ (W/mK) |
|---|---|---|---|---|---|---|
| Li | Li-1 | 288.20 | -7.2117 | 82.1294 | 2.7502 | 84.8796 |
|    | Li-2 | 293.08 | -7.6875 | 81.0457 | 2.6219 | 83.6676 |
|    | Li-3 | 303.64 | -7.7978 | 79.9341 | 2.6386 | 82.5727 |
|    | Li-4 | 305.35 | -7.9970 | 78.4404 | 2.5930 | 81.0334 |
|    | Li-5 | 313.87 | -8.1606 | 77.3224 | 2.6050 | 79.9274 |
|    | Li-6 | 316.28 | -8.2397 | 76.7194 | 2.5927 | 79.3121 |
| Be | Be-1 | 278.02 | -2.2042 | 220.9939 | 25.0928 | 246.0867 |
|    | Be-2 | 285.00 | -2.9479 | 216.3805 | 21.0501 | 237.4306 |
|    | Be-3 | 296.17 | -3.3623 | 196.5404 | 18.8916 | 215.4320 |
|    | Be-4 | 311.22 | -3.8448 | 174.4980 | 16.9305 | 191.4285 |
|    | Be-5 | 313.46 | -3.6524 | 164.2614 | 18.5753 | 182.8367 |
| Mg | Mg-1 | 290.95 | -4.8223 | 166.4749 | 2.1260 | 168.6009 |
|    | Mg-2 | 297.74 | -4.2645 | 148.2052 | 2.4020 | 150.6072 |
|    | Mg-3 | 302.19 | -3.6466 | 140.7474 | 2.3877 | 143.1351 |
|    | Mg-4 | 307.87 | -5.8340 | 140.4684 | 1.7672 | 142.2356 |
|    | Mg-5 | 311.41 | -5.1873 | 142.9120 | 2.0244 | 144.9364 |
| Al | Al-1 | 284.48 | -3.0485 | 232.7718 | 5.7530 | 238.5248 |
|    | Al-2 | 286.14 | -4.9667 | 231.2791 | 4.5353 | 235.8144 |
|    | Al-3 | 299.46 | -4.9448 | 223.9373 | 4.6760 | 228.6133 |
|    | Al-4 | 306.84 | -5.2123 | 223.6268 | 4.6600 | 228.2868 |
|    | Al-5 | 310.39 | -5.3456 | 220.0350 | 4.3882 | 224.4232 |
|    | Al-6 | 318.31 | -5.5119 | 209.5108 | 4.3312 | 213.8420 |
| Cu | Cu-1 | 285.31 | -7.2400 | 438.9252 | 2.0393 | 440.9645 |
|    | Cu-2 | 293.86 | -9.6536 | 399.0878 | 1.8816 | 400.9694 |
|    | Cu-3 | 298.49 | -8.7284 | 391.4515 | 2.1829 | 393.6344 |
|    | Cu-4 | 310.91 | -9.9209 | 374.4141 | 1.9523 | 376.3664 |

**5.3 Size effects of lattice thermal conductivity $\kappa_{ph}$ from NEAIMD**

It is well known that, in lattice dynamics each vibrational mode (phonon) has a specific wavelength. In view of this, finite size effects are inevitable in non-equilibrium molecular dynamics (NEMD) simulations of lattice thermal conductivity of most systems, where phonons are truncated due to the limited model length [13].

Due to the computational limitation of the AIMD, here we perform classical NEMD simulations of Cu at 300 K with different simulation cell length along the direction of heat transfer using the LAMMPS [14] package. The Cu-Cu interatomic interactions are described by the embedded-atom-method (EAM) potential [15]. In Fig. S6, we can see the strong finite size effect of $\kappa_{ph}$. The length of our NEAIMD model of Cu is 8 unit cells and our NEAIMD result of $\kappa_{ph}$ is about 2.18 W/mK around 300 K, which is similar to the classical NEMD results (3.84 W/mK) in Fig. S6.

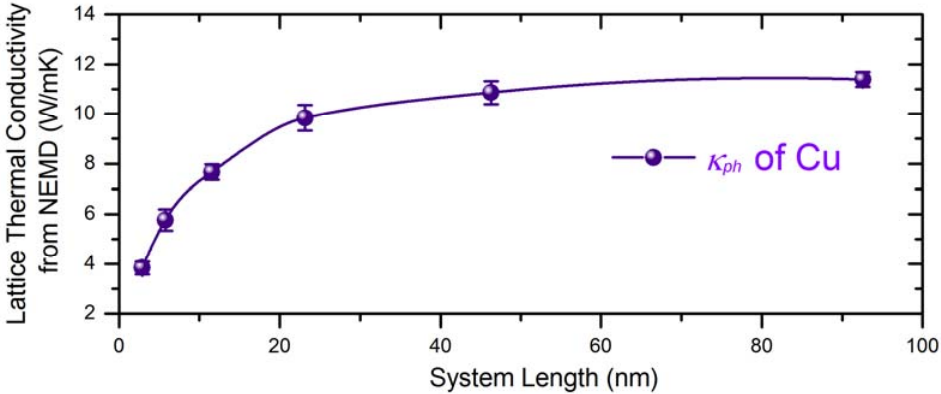

Fig. S6: Size effect of lattice thermal conductivity of Cu from non-equilibrium molecular dynamics simulation around 300 K.

To compare $\kappa_{ph}$ results from different calculation methods, we first calculate $\kappa_{ph}$ of different metals, at around 300 K, from classical equilibrium molecular dynamics (EMD) simulations, using the Green-Kubo method [13], as implemented in the LAMMPS package. The EAM potential parameters for Al [16], Cu [15], Be [17], Mg [18] and Li [19] are used to describe the interatomic interactions in the EMD simulations. The results are shown in Fig. S7.

Second, we calculate $\kappa_{ph}$ of metals at 300 K by solving the phonon Boltzmann transport equation (BTE), with force constants extracted from first-principles calculations. The phonon BTE model does not suffer from finite size effects. We employ the first-principle software package VASP [1,2] to calculate the second-order (harmonic) and third-order (anharmonic) force constants based on the finite displacement difference method [20, 21], and then use the ShengBTE package [21] to obtain $\kappa_{ph}$ by iteratively solving the BTE of phonons. The convergences of $\kappa_{ph}$ with respect to the $k$-grid size ($N \times N \times N$) in our calculations are fully examined and the parameter $N = 20$ is taken to evaluate the converged $\kappa_{ph}$. The convergences of $\kappa_{ph}$ with respect to the force cut-off distance are also examined and we took the distance of the fifth-order adjacent neighbor atoms as the force cut-off. The energy of plane-wave cutoff are adopted the value of 1.5 times of the default value in VASP pseudo-potential files.

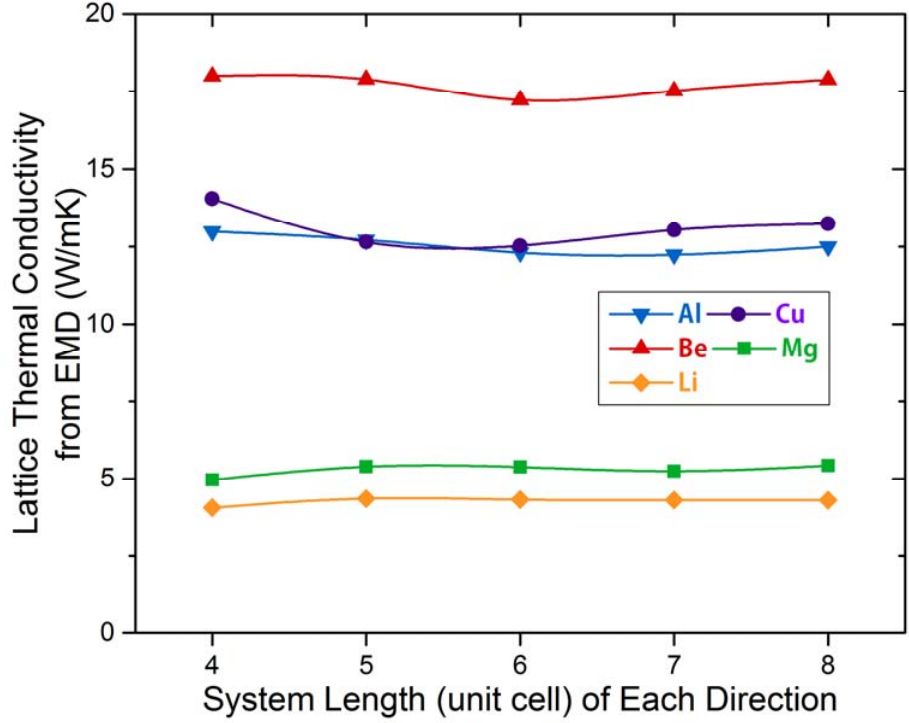

Fig. S7: Size effect of lattice thermal conductivity of metals from equilibrium molecular dynamics simulation around 300 K.

Finally, we compare $\kappa_{ph}$ from NEAIMD, classical EMD and phonon BTE method in Table 4. It can be seen that our NEAIMD simulations underestimate $\kappa_{ph}$ of Li, Al, Mg, and Cu. The effect is particularly significant for Cu. However, for Be, $\kappa_{ph}$ from NEAIMD is a little bit higher than the results from classical EMD and phonon BTE method.

We also examined the relationship between simulation size and non-linear effect for $\kappa_{el}$ from NEAIMD. The results can be found in Table 1 and Table 3. Unlike $\kappa_{ph}$, from our NEAIMD simulations results, we do not observe a clear size effect for $\kappa_{el}$.

Table 4. Comparison of lattice thermal conductivity $\kappa_{ph}$ (at approximately 300 K) calculated by NEAIMD, classical EMD and phonon BTE method.

| System | Li | Be | Mg | Al | Cu |
|---|---|---|---|---|---|
| NEAIMD $\kappa_{ph}$(W/mK) | 2.6386 | 18.8917 | 2.4021 | 4.6760 | 2.1829 |
| Classical EMD $\kappa_{ph}$(W/mK) | 4.3013 | 17.8805 | 5.4244 | 12.5067 | 13.2420 |
| BTE $\kappa_{ph}$(W/mK) | 3.2080 | 15.2697 | 8.3925 | 6.3789 | 20.4131 |

## 6. Analysis of propagation of errors

From the knowledge of the statistical average propagation of errors [22], we know that, when a variable is defined as $X = \frac{u}{v}$, the square of the error in $X$ can be expressed as

$$\sigma_X^2 = \sigma_{\frac{u}{v}}^2 = \sigma_u^2 \left(\frac{\partial X}{\partial u}\right)_{\bar{u}}^2 + \sigma_v^2 \left(\frac{\partial X}{\partial v}\right)_{\bar{v}}^2 = \frac{\sigma_u^2}{\bar{v}^2} + \frac{\sigma_v^2 \bar{u}^2}{\bar{v}^4}$$

$$\Rightarrow \left(\frac{\sigma_X}{\frac{\bar{u}}{\bar{v}}}\right)^2 = \left(\frac{\sigma_{\frac{u}{v}}}{\frac{\bar{u}}{\bar{v}}}\right)^2 = \left(\frac{\sigma_u}{\bar{u}}\right)^2 + \left(\frac{\sigma_v}{\bar{v}}\right)^2. \qquad (19)$$

From Equations (17) and (19), we know that the error in $\kappa_{el}$ mainly originates from $\nabla T$ and $\frac{\partial \bar{U}_{EPO}(l)}{\partial N_l}$. As the $\nabla T$ calculation is based on the statistical time average of temperature of each single atom layer, the temperature fluctuation $\overline{(\Delta T)^2} = k_B T^2 / C_v$ [23] of each layer is large due to the small number of atoms in the layer. Thus, the conventional error estimate for $\nabla T$ is quite large. However, from Fig. S2 we find that the NEAIMD always yields a stable temperature profile after sufficient simulation time, and so we have reason enough to assume the linear fitting error as the error in $\nabla T$.

At the same time, we notice that the non-linear phenomenon of $\frac{\partial \bar{U}_{EPO}(l)}{\partial N_l}$, which was discussed in Sec. 4.2, leads to a relatively large error in $\kappa_{el}$. We calculate the error in $\kappa_{el}$ (Table 5) from Equation (19). From Table 5 we can see that the $\kappa_{el}$ of Be and Al have relatively large uncertainties, because of the large error in $\frac{\partial \bar{U}_{EPO}(l)}{\partial N_l}$.

Table 5. Error bar of electrical thermal conductivity $\kappa_{el}$ of metals around 300 K.

| System | Li | Be | Mg | Al | Cu |
|---|---|---|---|---|---|
| Temperature (K) | 303.64 | 296.17 | 297.74 | 299.46 | 298.49 |
| Error bar of linear fitting $\nabla T$ | 4.06% | 10.36% | 2.86% | 8.57% | 4.74% |
| Error bar of linear fitting $\frac{\partial \bar{U}_{EPO}(l)}{\partial N_l}$ | 7.15% | 27.04% | 15.88% | 28.61% | 7.19% |
| Total error bar of $\kappa_{el}$ (W/mK) | 6.79 | 62.38 | 24.30 | 67.90 | 33.88 |
| Total error bar of $\kappa_{el}$ (percentage) | 8.22% | 28.96% | 16.13% | 29.86% | 8.61% |

## 7. Comparing our results with conventional BTE method and experimental data

In order to examine our theory and evaluate the results of our method, we compare our simulation results of some common metals with experimental measurements and the results from the traditional BTE framework.

In Sec. 5.3, we calculated the lattice thermal conductivity, $\kappa_{ph}$, of metals with the BTE method based on first-principle calculations. From the free electron model and BTE theory, we can estimate the electronic thermal conductivity $\kappa_{el}$ as [11]

$$\kappa_{el} = \frac{\pi^2 n k_B^2 T \tau_{el}}{3m}, \quad (20)$$

where $n$ is the concentration of free electrons, $m$ is the electron mass, $k_B$ is Boltzmann constant, $T$ is system temperature, and $\tau_{el}$ is the collision time of free electrons. The value of $\frac{\kappa_{el}}{\tau_{el}}$ is calculated using the BoltzTraP package [24] to solve BTE for electrons based on the electronic band structures, which are calculated by VASP. As in usual practice, we take $\tau_{el} = 1 \times 10^{-14}$ s [24, 25], allowing us to obtain $\kappa_{el}$ from the conventional BTE framework.

Finally, we get the total thermal conductivity of metals ($\kappa$) by summing up the lattice thermal conductivity ($\kappa_{ph}$) and electronic thermal conductivity ($\kappa_{el}$). The results of $\kappa_{ph}$, $\kappa_{el}$, and $\kappa$ from different methods along with the experimental measurement data are reported in Table 6. The comparison clearly shows that our method is much better at predicting the thermal conductivity of metals than the conventional BTE method.

Table 6. Comparison of $\kappa_{ph}$, $\kappa_{el}$, and $\kappa$ of simulated metals among our NEAIMD method, conventional BTE method, and experiments at 300 K.

| System | NEAIMD $\kappa_{el}$ (W/mK) | BTE $\kappa_{el}$ (W/mK) | NEAIMD $\kappa_{ph}$ (W/mK) | BTE $\kappa_{ph}$ (W/mK) | NEAIMD total $\kappa$ (W/mK) | BTE total $\kappa$ (W/mK) | Exp. $\kappa$ (W/mK) |
|---|---|---|---|---|---|---|---|
| Li | 79.9341 | 70.3525 | 2.6386 | 3.2080 | 82.5727 | 73.5605 | 84.8 |
| Be | 196.5404 | 48.1385 | 18.8917 | 15.2697 | 215.4321 | 63.4082 | 200 |
| Mg | 148.2052 | 105.4915 | 2.4021 | 8.3925 | 150.6073 | 113.8840 | 156 |
| Al | 223.9373 | 221.5790 | 4.6760 | 6.3789 | 228.6133 | 227.9579 | 237 |
| Cu | 391.4515 | 136.0173 | 2.1829 | 20.4131 | 393.6344 | 156.4304 | 401 |

8. **The exponential autocorrelation time of velocity of EPO at ion core**

In calculation of the spectral density of electrostatic potential oscillating velocity ($S_{\Delta U}$) in Sec. 2, after calculating the autocorrelation function of velocity of EPO at ion core $\Delta U_{ion}(t)$, we performed an exponential decay fitting of the autocorrelation function with the formula

$$y = A\exp(-t/\tau_{\exp}), \quad (21)$$

where $\tau_{\exp}$ is called the exponential autocorrelation time. It is surprising to find that $\tau_{\exp}$ is approximately on the same order of magnitude as the collision time of free electrons ($\tau_{el} = 1 \times 10^{-14}$ s). We examined all cases of metals and present the data of $\tau_{\exp}$ in Table 7. The order of magnitude of our results ($\sim 10^{-14}$ s) agrees very well with the common theoretical value [24-26]. We anticipate that there must be some physical mechanisms behind this phenomenon.

Table 7. Exponential autocorrelation time $\tau_{exp}$ of velocity of EPO at ion core (300 K)

| System | Li | Be | Mg | Al | Cu |
|---|---|---|---|---|---|
| Temperature (K) | 302.32 | 300.67 | 299.22 | 299.89 | 300.72 |
| $\boldsymbol{\tau_{exp}}$ ($\times 10^{-14} s$) | 1.258 | 0.584 | 1.617 | 1.282 | 1.431 |

## 9. Advantages, limitations, challenges, and future work of this method
### 9.1 Advantages

1) Our NEAIMD-EPO method provides a direct and clear procedure for simulating the thermal transport behavior of free electrons from an atomistic point of view. It will be very advantageous for investigations of very-large-scale integration (VLSI), of relevance to the semiconductor industry.
2) The NEAIMD-EPO method naturally but implicitly includes the complicated interactions between electrons and electron-phonon coupling.
3) The NEAIMD-EPO approach is a new method which can calculate electronic thermal conductivity without artificial manipulation and input parameters.
4) The NEAIMD-EPO framework also provides the physical picture of how the thermal energy is carried by thermally excited electrons and how this energy is transported in metals.

### 9.2 Challenges

1) The nonlinear phenomenon of the effective amplitude of EPO along the heat flux direction in some metals still needs further study.
2) A coherent understanding why the exponential autocorrelation time of velocity of EPO at ion cores has the same order of magnitude as the collision time of free electrons is still missing.

### 9.3 Limitations

1) As our NEAIMD-EPO framework is built on the free electron gas model, so far, this method is limited to simulation of pure metals.
2) So far, this method cannot be directly used to simulate thermal transport of metals at low temperatures.
3) As this method is realized in the ab initio molecular dynamics simulation, the simulation results will depend on the pseudopotential used.
4) The computation costs for the NEAIMD simulations are much higher than that of normal density functional theory (DFT) simulations.

### 9.4 Future work

With the theory and computational capacity improving, the NEAIMD-EPO method shows the potential in investigating alloys, semiconductors, metal/non-metal interfaces, and even directly simulating nano-devices in the future. It will be promising in theoretical study of the nanotechnology.



\end{document}